\documentclass[10pt]{iopart}
\usepackage{graphicx}
\begin{document}
\title[Electron dynamics at a positive ion]{Nonlinear response of electron dynamics to a positive ion}
\author{James W. Dufty\dag\ \footnote[3]{To whom correspondence should be addressed (dufty@phys.ufl.edu)},
Ilya V. Pogorelov\dag\, Bernard Talin\ddag, and Annette Calisti\ddag}
\address{\dag\ Department of Physics, University of Florida, Gainesville, FL 32611, USA}
\address{\ddag\ Universit\'{e} de Provence, CNRS UMR 6633, Centre Saint J\'{e}r\^{o}me, 13397 Marseille Cedex 20, France}

\begin{abstract}
Electric field dynamics at a positive ion imbedded in an electron
gas is considered using a semiclassical description. The
dependence of the field autocorrelation function on charge number
is studied for strong ion-electron coupling via MD simulation. The
qualitative features for larger charge numbers are a decreasing
correlation time followed by an increasing anticorrelation.
Stopping power and related transport coefficients determined by
the time integral of this correlation function result from the
competing effects of increasing initial correlations and
decreasing dynamical correlations. An interpretation of the MD
results is obtained from an effective single particle model
showing good agreement with the simulation results.
\end{abstract}


\maketitle
\section{Introduction}
The total electric field at a particle in a plasma determines the
dominant radiative and transport properties of that particle. The
properties of fields due to positive ions at a positive particle
have been studied in some detail for both the static distribution
of fields \cite{Duftymicro} and the dynamics of the electric field
autocorrelation function \cite{Boercker,Berkovsky}. The latter
poses a real challenge since finite charge on the site at which
the field is considered precludes the use of standard linear
response methods. The corresponding study of fields at a positive
ion due to electrons has been considered more recently for the
simplest case of a single ion of charge number $Z$ in an
semiclassical electron gas. The static properties (electron charge
density, electron microfield distribution) have been discussed in
some detail elsewhere \cite{Talin}. Here, attention is focused on
the dynamics via the electric field autocorrelation function. The
case of opposite sign electron fields at a positive ion is
qualitatively different from same sign ion fields, since in
the former case electrons are attracted to the ion leading to
strong electron-ion coupling for the enhanced close
configurations. It is difficult {\it a priori} to predict the
qualitative features of the correlation function due to this
inherent strong coupling. Consequently, the analysis here has been
based on MD simulation of the correlation function followed by an
attempt to model and interpret the observed results. The
simulations represent classical mechanics for Coulomb interactions
with the ion-electron potential modified at short distances to
represent quantum diffraction effects. The details have been
discussed elsewhere \cite{Talin} and will not be repeated here.
There are only three dimensionless parameters: the charge number
of the ion, $Z$, the electron-electron coupling constant $\Gamma
$, and the de Broglie wavelength relative to the interelectron
distance, $\delta $. The electron-ion coupling is measured by the
maximum value of the magnitude of the regularized ion-electron potential at the
origin, $\sigma =Z\Gamma /\delta $. In this brief report results
are reported for $\Gamma =0.1$ and $\sigma =0.25Z$, with
$Z=8,20,30,$ and $40$. The corresponding density and temperature
are $n=2.5\times 10^{22}$ cm$^{-3}$and $T=7.9\times 10^{5}$
$K$. 

The primary observations from the simulations of the
field autocorrelation function for increasing charge number are:
1) an increase in the initial value, 2) a decrease in the
correlation time, and 3) an increasing anticorrelation at longer
times. The stopping power for the ion by the electron gas is
proportional to the time integral of the field autocorrelation
function in the limit of large ion mass \cite{Berkovsky2}. This
same integral determines the self-diffusion and friction
coefficients in this same limit \cite{Berkovsky3}. The nonlinear
dependence of these properties on $Z$ is therefore the result of
competition between the increase of the integral due to 1) and the decrease 
due to 2) and
3). The results from simulation show that the latter two dynamical
effects dominate the former static effect. To interpret this, a
simple model for the field correlation function is proposed such
that the initial correlations are given exactly, but the dynamics
is determined approximately from a single electron-ion trajectory.
The model reproduces well the MD simulation results and suggests
an interpretation of 2) and 3).

\section{MD simulation of field dynamics}
The system considered consists of $N_{e}$ electrons with charge
$-e$, an infinitely massive positive ion with charge $Ze$ placed
at the origin, and a rigid uniform positive background for overall
charge neutrality. The regularized electron-ion potential is
chosen to be $V(r)=-Ze^{2}\left( 1-e^{-r/\delta }\right) /r$ \ $\
$where $\delta =\left( 2\pi \hbar ^{2}/m_{e}k_{B}T\right) ^{1/2}$
is the electron thermal de Broglie wavelength. For values of
$r>>\delta $ the potential becomes Coulomb, while for $r<<\delta $
the Coulomb singularity is removed and $V(r)\rightarrow
-Ze^{2}/\delta$. This is the simplest phenomenological form
representing the short range effects of the uncertainty principle
\cite{Minoo}. Dimensionless variables are based on scaling
coordinates with the average electron-electron distance, $r_{0}$,
defined in terms of the electron density $n_{e}$ by $4\pi
n_{e}r_{0}^{3}/3=1$, and scaling time with the inverse electron
plasma frequency. The electron electric field at the ion is
obtained from the total regularized potential
\begin{equation}
\mathbf{E}=-\nabla _{0}V(\left\{ r_{i0}\right\}
)=\sum_{i=1}^{N}\mathbf{e} \left( \mathbf{r}_{i0}\right)
\label{2.4}
\end{equation}
where $\mathbf{r}_{i0}=\mathbf{r}_{i}-\mathbf{r}_{0}$ is the
position of the $i^{th}$ electron relative to the ion, and
\begin{equation}
V(\left\{ r_{i0}\right\}
)=\sum_{i=1}^{N}V(r_{i0}),\hspace{0.4cm}\mathbf{e} \left(
\mathbf{r}_{i0}\right)=e\frac{\widehat{\mathbf{r}}_{i0}}{r_{i0}^{2}}
(1-\left( 1+\frac{r_{i0}}{\delta }\right) e^{-r_{i0}/\delta })
\label{2.5}
\end{equation}
The dimensionless field autocorrelation function is defined by
\begin{equation}
C(t)=\frac{r_{0}^{4}}{e^{2}}<\mathbf{E}\left( t\right)
\mathbf{\cdot E}>. \label{2.6}
\end{equation}
The brackets denote an average over the classical Gibbs ensemble
for the composite electron-ion system at equilibrium. 

\begin{figure}
\begin{center}
\includegraphics[width=10cm]{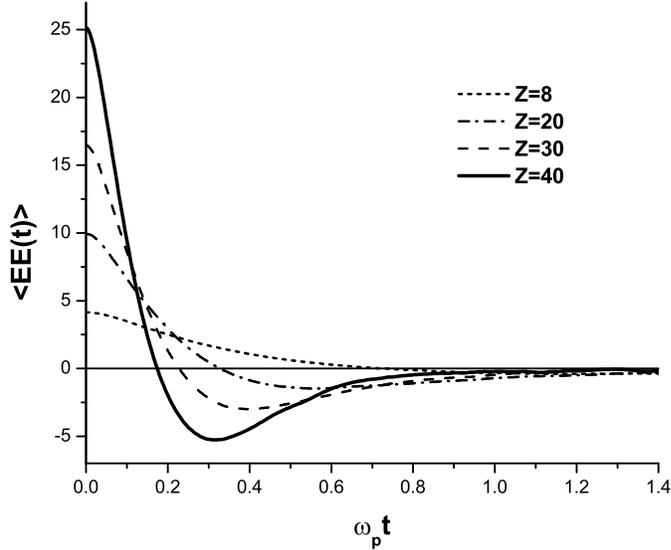}
\end{center}
\caption{\label{Figure 1}Field autocorrelation function for $Z= 8,20,30,40$ at $\Gamma =0.1$ and $\delta = 0.4$.}
\end{figure}

The results for $C(t)$ from MD simulation are shown in Figure 1 for
$Z=8,20,30,$ and $40$. The initial value increases approximately
as a third order polynomial in $Z$ \cite{Talin}. The correlation
time $t_{c}$ is defined to be the time at which the correlation
function first goes to zero, $C(t_{c})=0$, and is seen to decrease
as $Z$ increases. Finally, in all cases there is anticorrelation
($C(t)<0$) for $t>t_{c}$. The physical basis for the increase in
the initial value is easy to understand. As $Z$ increases, the
electron density near the ion increases and the magnitude of the
field increases for these more probable closer configurations. An
explanation for the correlation time and anticorrelation is more
difficult, and is the objective of the following sections. First,
some consequences of this behavior are illustrated.

\section{Stopping power, friction, and self-diffusion}
The case of an infinitely massive ion considered here leads to
exact relationships between transport coefficients characterizing
three physically different phenomena: 1) the low velocity stopping
power $\mathcal{S}$ for a particle injected in the electron gas,
2) the friction coefficient $\xi $ for the resistence to a
particle being pulled through the gas, and 3) the self-diffusion
coefficient $D$ of a particle at equilibrium with the gas. The
exact relationship is \cite{Berkovsky3}
\begin{equation}
m_{0}\xi =\left( \beta D\right)
^{-1}=\mathcal{S}(v)/v=\beta Z^{2}r_{0}^{-4} \int_{0}^{\infty }dtC(t)
\label{3.1}
\end{equation}
This Green-Kubo representation allows determination of these
transport properties from an equilibrium MD simulation. Previous
simulations of stopping power have studied the nonequilibrium
state of the injected particle, measuring directly the energy
degradation \cite{Zwick}. As discussed below, (\ref{3.1}) provides
the basis for additional interpretation of the results. 

\begin{figure}
\begin{center}
\includegraphics[width=10cm]{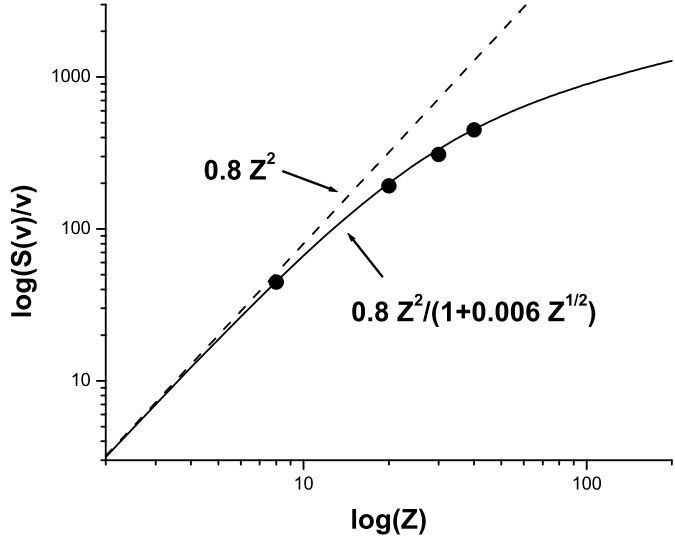}
\end{center}
\caption{\label{Figure 2}Stopping power $\mathcal{S}(v)/v$ for $Z= 8,20,30,40$ at $\Gamma =0.1$ and $\delta = 0.4$.}
\end{figure}

Figure 2 shows the dimensionless stopping power obtained from the results
of Figure 1 as a function of $Z$. Also shown is the Born
approximation $\propto Z^{2}$, valid for small $Z$. Previous
simulations \cite{Zwick} and some experiments \cite{Experiment}
suggest a crossover at larger $Z$ to a weaker growth $\propto
Z^{2}$. The data in Figure 2 has been fit to a such a crossover
function showing consistency with these earlier results. This
behavior is somewhat puzzling in light of the strong growth of the
initial value $C(0)\approx Z^{3}$ at large $Z$. Making this
explicit, the stopping power can be expressed as

\begin{equation}
\mathcal{S}(v)/v\propto Z^{5}\int_{0}^{\infty
}dtf(t),\hspace{0in}\hspace{0.5in}f(t)=C(t)/C(0)  \label{3.2}
\end{equation}
Evidently, the dynamical effects of the normalized correlation
function $f(t) $ \emph{decrease} the stopping power as $\sim
Z^{-3.5}$ for large $Z$. A possible explanation for this is given
in the following section.

\section{Single particle dynamics}
Consider again the initial covariance $C(0)$ for which three
representations can be given
\begin{eqnarray}
C(0) &=&\frac{3}{4\pi }\int d\mathbf{re}\left( \mathbf{r}\right)
\mathbf{\cdot }\left[ g_{ie}(r)\mathbf{e}\left( \mathbf{r}\right)
+\frac{3}{4\pi}\int d\mathbf{r}^{\prime
}g_{ie}(\mathbf{r},\mathbf{r}^{\prime })\mathbf{e}
\left( \mathbf{r}^{\prime }\right) \right]   \nonumber \\
&=&\frac{3}{4\pi Z\Gamma }\int d\mathbf{r}g_{ie}(r)\nabla
\mathbf{\cdot e}\left( \mathbf{r}\right)   \nonumber \\
&=&\frac{3}{4\pi }\int d\mathbf{r}g_{ie}(r)\mathbf{e}_{mf}\left(
\mathbf{r} \right) \mathbf{\cdot e}\left( \mathbf{r}\right)
\label{3.4}
\end{eqnarray}
The first equality expresses the covariance in terms of the one
and two electron correlations with the ion, $g_{ie}(r)$ and
$g_{ie}(\mathbf{r}, \mathbf{r}^{\prime })$ respectively. The
second equality exploits the relationship for the field to the
Gibbs factor $\beta U(\left\{ r_{i0}\right\} )=Z\Gamma V(\left\{
r_{i0}\right\} )$ and an integration by parts. This second
representation requires only the one electron correlation
function. The third representation is obtained from the second by
an integration by parts to identify the mean force field
$\mathbf{e}_{mf}\left( \mathbf{r}\right) =\nabla \ln g_{ie}(r)$.
The third equality of (\ref{3.4}) is similar to the first with apparent
neglect of the two electron-ion correlations. However, these
latter contributions are incorporated exactly in the mean force
field $\mathbf{e}_{mf}\left( \mathbf{r} \right) $. This suggests a
corresponding model for finite times

\begin{equation}
C(t)\rightarrow \frac{3}{4\pi }\int d\mathbf{v}d\mathbf{r}\phi (v)g_{ie}(r)\mathbf{e}
_{mf}\left( \mathbf{r}\right) \mathbf{\cdot e}\left(
\mathbf{r} (t)\right) \label{3.6}
\end{equation}
where $\phi(v)$ is the normalized Maxwellian and $\mathbf{r}(t)$ is a single electron
trajectory in the presence of the ion, generated by the potential
associated with $g_{ie}(r)$ for stationarity. Further details and a
justification based on the Vlasov equation will be given
elsewhere. Clearly, the initial covariance is given exactly by
this approximation. For practical purposes $g_{ie}(r)$ and the
associated field and potential are represented here by the
non-linear Debye-Huckel approximation with effective charge number
and screening length adjusted to fit the MD data.

\begin{figure}
\begin{center}
\includegraphics[width=10cm]{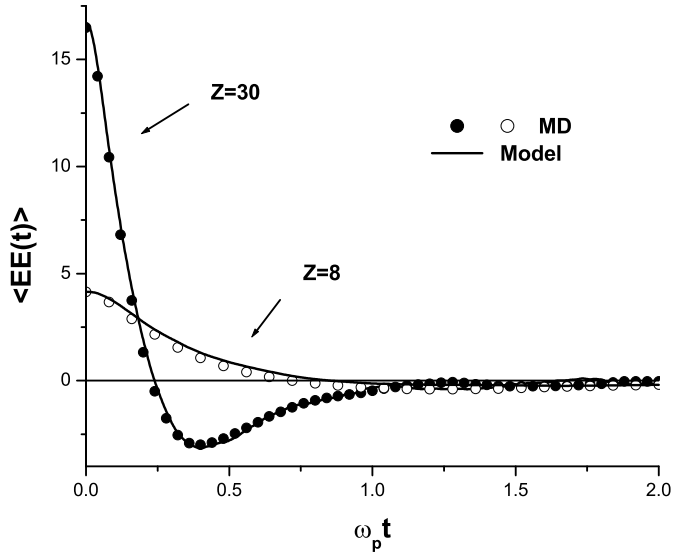}
\end{center}
\caption{\label{Figure 3}Comparison of $C(t)$ from MD and from Eq.\ref{3.6}
  for $Z= 8$ and $30$ at $\Gamma =0.1$ and $\delta = 0.4$.}
\end{figure}

Figure 3 shows a comparison of (\ref{3.6}) with the MD results of
Fig. 1 for $Z=8$ and $Z=30$. The agreement is very good, indicating that the dominant
$Z$ dependence is captured by the initial correlations and the
single particle dynamics.

\section{Discussion}
The good agreement of the simple model in the previous section
suggests that the decrease  in correlation time and build up of
anticorrelation can be understood in terms of one electron
dynamics. As $Z$ increases the probability of close electron ion
configurations increase. The electron is subjected to greater
acceleration toward the ion and the time to reverse its sign
decreases. This is the effect of decreasing correlation time. Once
the field has reversed sign there is anticorrelation. Since the
closer configurations imply larger field values the magnitude of
this anticorrelation also increases with increasing $Z$. It
remains to quantify this picture but the  single electron dynamics
appear to provide qualitative confirmation. A more complete
discussion will be provided elsewhere.

\section{Acknowledgments}

Support for this research has been provided by the U.S. Department
of Energy Grant No. DE-FG03-98DP00218. J. Dufty is grateful for the
support and hospitality of the University of Provence.


\begin{thebibliography}{9}
\bibitem{Duftymicro} Dufty J W in {\it Strongly Coupled Plasmas}, 493,  de
Witt H and Rogers F editors (NATO ASI Series, Plenum, NY, 1987).
\bibitem{Boercker} Boercker D, Iglesias C, and Dufty J W 1987 {\it Phys.
Rev.} A {\bf 36}, 2254
\bibitem{Berkovsky} Berkovsky M, Dufty J W, Calisti A, Stamm R, and Talin B 1996
{\it Phys. Rev. E} {\bf 54}, 4087-4097.
\bibitem{Talin} Talin B, Calisti A and Dufty J W 2002 {\it Phys.
Rev.} E {\bf 65}, 056406
\bibitem{Berkovsky2} Dufty J W and Berkovsky M 1995 {\it Nucl. Inst. Meth.} B {\bf 96}
626\bibitem{Berkovsky3} Dufty J W and Berkovsky M in {\it Physics of
Strongly Coupled Plasmas} edited by Kraeft W, Schlanges M,
Haberland H and Bornath T (World Scientific, River Edge, NJ, 1996)
\bibitem{Minoo} Minoo H, Gombert M and Deutsch C 1981 {\it Phys. Rev.} A {\bf 23} 924
\bibitem{Zwick} Zwicknagel G, Toepffer C and Reinhard P-G 1999 {\it Physics
Reports} {\bf 309} 118
\bibitem{Experiment} Winkler Th. et al.  1997 {\it Nucl. Inst. and
Meth.} A {\bf 391} 12
\end{thebibliography}
\end{document}